# On Performance of Logical-Clustering of Flow-Sensors

Rahim Rahmani, Hasibur Rahman and Theo Kanter

Department of Computer and Systems Sciences, Stockholm University
Kista, Stockholm SE-164 40, Sweden

**Abstract**
In state-of-the-art Pervasive Computing, it is envisioned that unlimited access to information will be facilitated for anyone and anything. Wireless sensor networks will play a pivotal role in the stated vision. This reflects the phenomena where any situation can be sensed and analyzed anywhere. It makes heterogeneous context ubiquitous. Clustering context is one of the techniques to manage ubiquitous context information efficiently to maximize its potential. Logical-clustering is useful to share real-time context where sensors are physically distributed but logically clustered. This paper investigates the network performance of logical-clustering based on ns-3 simulations. In particular reliability, scalability, and reachability in terms of delay, jitter, and packet loss for the logically clustered network have been investigated. The performance study shows that jitter demonstrates 40 % and 44 % fluctuation for 200 % increase in the node per cluster and 100 % increase in the cluster size respectively. Packet loss exhibits only 18 % increase for 83 % increase in the packet flow-rate.

## 1. Introduction

The Wireless Sensor Networks (WSNs) is an integral part of today's pervasive computing and expected to play a pivotal role in the future Networked Society. The primary use of sensors is to collect data from physical objects. It is foreseen that any situation can be sensed and analyzed anywhere which leads to more and more sensors deployment in today's Internet infrastructure and sensors are made available to the services through the distributed acquisition and dissemination of sensor data assembled from physical objects. Services can access this heterogeneous context information anywhere. The use of sensors is increasing rapidly. Billions of sensors will be used in the foreseeable future [2]. This will play a vital role in making context information accessible for anyone and anything in the future Networked Society [1]. These enormous numbers of sensors deployment in the Internet of Things (IoT) will allow gathering information from people, places, and objects i.e. from distributed sensor networks. Spontaneous human participation which is known as crowdsourcing is also envisioned [6]. This implies that rapid real-time data will be generated by crowd about the circumstances surround [7]. These will produce heterogeneous context information. Moreover, a single sensor might produce different data. For example, sensors carried by human on their smart devices might produce different data in different time. This necessitates proper management of heterogeneous contexts obtained from sensors. Data management should be reliable, and the high volume of data should be scaled appropriately in order to use efficiently and meaningfully. Clustering the context i.e. data is one of the proficient applications. Furthermore, it will be advantageous to cluster sensor data based on context similarity. In one of our previous papers, logical-clustering of flow-sensors has been presented [2]. Logical-clustering implies that sensors might reside remotely physically but clustered logically based on context similarity. Previous most work on sensors clustering concentrated on physical location nearness for energy and routing management, and to increase system scalability and robustness. Context in sensors clustering has been discussed too, but in all cases definition of context is specific. Moreover, their solution is limited to neighboring sensors. However, the concept of logical-clustering will allow resources (data, services) to be shared among different physically distributed sensors in distributed sensor networks. Sensors can share resources through distributed collaboration which was lacking in the existing management of context information. Once the clustering is done then each cluster is identified through a *context-ID* which is defined based on context similarity and published on the internet. Any interested sensor, may be located remotely, can subscribe to the context-ID.

OpenFlow based sensors are known as flow-sensors [3]. It has been proven that flow-sensors perform better than typical sensors [3]. However, it will be infeasible for a single OpenFlow controller to manage the increased number of sensors. In order to manage huge amount of sensors, more than one OpenFlow controller is desirable. HyperFlow addresses the issue and offers multiple controllers which are physically distributed but logically centralized [2]. The controllers are synchronized and can be resilient for network slicing. An important factor that was missing in the existing OpenFlow specification is interconnection between different OpenFlow networks, HyperFlow solves this problem by using the publish/subscribe mechanism. In HyperFlow, each controller can make decision locally which minimizes the response time. Controllers exchange messages to notify about any network-wide changes. These logically synchronized controllers are called *logical-sink* [2].

Network performance is one of the most researched issues in the field of wireless sensor networks. Network management becomes an important consideration as the number of sensor nodes increases. In future, network will encounter thousand times traffic volumes compared to today's traffic volumes. Latency, reliability, scalability, and data reachability are few of the challenges that future network would encounter [1]. Therefore, it is essential to design network carefully so that network does not incur performance degradation. In our previous paper, the feasibility and technical presentation of logical-clustering have been discussed [2]. In addition, computational efficiency of logical-clustering has also been shown in [2]. In this paper, the focus will be on investigating few of the significant network performance metrics of logical-clustering. The network has been designed in ns-3 (network simulator). A performance study has been made in terms of delay, jitter and packet loss to verify the reliability, scalability and reachability of the designed network. Hence, the main focus of this paper will be to:

- Design a WSN of logical-clustering of flow-sensors in ns-3
- Verify the reliability of the designed network in terms of packet delay and jitter
- Verify packet reachability
- Examine scalability of the network for increased number of nodes and groups
- Provide use cases of logical-clustering

The remainder of the paper is organized as follows: section 2 presents the related work. Section 3 discusses the motivation behind the work. Section 4 outlines the system model considered for the proposal. Next, section 5 describes the model checking of the proposal. Simulation results are analyzed in section 6 which is followed by section 7 that illustrates few of the possible use cases of the proposed concept. Finally section 8 concludes the work and a guideline for future work is presented.

## 2. Related Work

There have been many researches about clustering in the WSNs. Most of the previous researches have been on preserving energy and prolonging the battery for the re-source-constrained sensor nodes. For example, LEACH [4] is the first clustering technique for achieving network longevity and energy dissemination reduction. Padmanabhan and Kamalakkannan in [5] further modified LEACH to prolong the network stability. Kumar et al. in [27] also examined different LEACH techniques in a view to prolong network lifespan. Literatures in [8 – 10, 24] discuss clustering of sensors but for the sake of data-accumulation. Clustering of sensors helps in reducing energy consumption, stabilizing network, efficient routing etc. S. Bandyopadhyay et al. in [8] analyzed hierarchical clustering and discussed that energy consumption is decreased if clustering level of hierarchy is increased. Abbasi, Younis and Lotfinezhad, Liang in [9-10] mention that inter-cluster communication is only limited to cluster-heads which results in communication bandwidth saving and in reducing message exchanges between sensors. Hyun and Hyuk in [28] discussed that efficient cluster-head selection prolongs the network life span and saves energy. D. Ma et al. in [25] proposed a clustering protocol with dual cluster-head concept to further improve network life time and more data accumulation to the base station.

Lombriser et al. in [11] presented distributed processing of context for dynamic WSNs. Their proposed E-SENSE computes context information from sensor networks. Sensors are clustered based on context-activity but limited only to neighboring sensors. It does not solve large-scale sensor network issue. Franco in [6] envisioned the idea of sensing, actuating and computing of anything anywhere for the future pervasive computing. He further outlined that spontaneous human participation i.e. crowdsourcing is vital for distributed collaboration to enrich urban networks. G. Barbier et al. in [7] presented maximizing the data obtained through crowdsourcing. They portrayed that crowdsourcing is faster and beneficial. With crowdsourcing, any event can be detected and analyzed. Event in the urban areas are fast changing. Moreover, some events are recursive and some are non-recursive [12]. Scalability, reachability and reliability of the obtained data from urban events through crowdsourcing become a challenge. Guo and Han in [13] discussed the reliability issue in data collection for WSN. They discussed the essence of reliable data collection for mobile nodes. The importance of latency in reliability for mobile WSNs has been discussed by Y. Rao et al. in [26]. They proposed a clustering based routing protocol for reliable data packet delivery in real-time. Ericsson in [1] further outlined the significance of reliability, latency, delay, maximum service (data) delivery i.e. reachability etc. for the future Networked Society.

Luca and Gian in [14] introduced logical-neighborhood of sensor nodes which replaced physical neighborhood concept. This idea more or less resembles our proposal. However, their solution is a programming language abstraction where nodes are said to be in the logical neighborhood if certain attributes are satisfied. A programmer defines the nodes' attributes and the data segment that can be part of a neighborhood. Therefore, it does not explicitly solve the real-time context sharing issue which is the prime objective of our proposal. In this paper, our focus is to examine the network performance of the logical-clustering of flow-sensors.

## 3. Motivation

Traditionally, sensors are used to obtain data from physical objects. Sensors also collaborate to achieve common goals. With the technological advancement, sensing devices have become more intelligent and affordable. Hence, the applicability of sensors is always rising, and it is believed that billions of sensors will be deployed in the future. Moreover, sensors are fundamental in the Internet of Things (IoT) deployment for any kind of urban event detection. These are used for different purposes and to obtain heterogeneous data from distributed sensor networks. Sensors deployment can be both deterministic (fixed) and random (mobile). Therefore, real-time context sharing will be a big challenge to existing and later distributed WSNs applications. Earlier solutions do not provide proper management of context information; hence current context information management does not support real-time sharing of context and do not scale well for heterogeneous interoperability. This necessitates proper management of the obtained data i.e. context information from sensors in order to use in an efficient and useful way. Most researches thus far concentrated on decreasing energy consumption so that sensors longevity is ensured. Several researchers have worked on clustering sensor nodes too, but again largely for sensor nodes stability. There have been some proposals for sensors data-management, but their proposals restrict to a certain area for adjacent sensors. It is also important that context generated by the sensors should be used meaningfully to take its full advantages. Real-time context sharing will be beneficial when clustered based on context similarity. The idea of clustering the sensors logically based on context similarity would allow resources (data, services) to be shared. Furthermore, the idea will provide topological sensor networks with scalability, reliability and high reachability in terms of delay, latency and packet loss.

## 4. System Model

Some of the definitions that have been used for modeling the system are presented below.

Sensor-ID: Sensors should have unique IDs. A *sensor-ID* can be obtained in different ways, e.g. the ID can be chosen randomly or can be obtained by hashing the sensor IP or MAC address [15].

Flow-ID: *Flow-ID* is the logical identification of the flow. According to [16], a flow could be defined based on capabilities of a particular implementation. The flow-ID is the flow packets from a particular sensor to the sink. As long as the sensor is interested in the same flow packets, the flow-ID remains same. But if sensor changes the flow of packets, the flow-ID is also changed. OpenFlow flow-tables consist of match-fields (i.e. packet header), action sets and statistics. The packet header defines the flow and action defines the flow-ID.

Context-ID: The *context-ID* is the identifier of a cluster. This can be compared to the idea that of a hashtag. As hashtag groups the similar messages, context-ID has the same objective. Context-ID is a mean of clustering similar data. The context-ID is published to the internet through the logical-sink and any interested entity i.e. sensor can subscribe to the context-ID.

Context flow-table: OpenFlow specification implies that match fields can be defined according to the research requirement [16]. A new flow-table for flow-sensor has been defined which includes flow-sensor's sensor-ID, flow-ID and context-ID. This flow-table is named *context flow-table*.

### 4.1 Network

A two-tier H-DHT system model has been considered. Controlling the ever-increasing number of sensors would be infeasible for single logically centralized controller (current OpenFlow standard), and in order to scale well for enormous number of sensors, the idea of HyperFlow (HF) has been exploited. This implies that multiple numbers of controllers (sinks) in the network has been used. The sinks are physically distributed but logically synchronized, hence this idea has been defined as *logical-sink* [2]. Another advantage of utilizing logical-sink is that each sink can do processing locally. And then other sinks get notified of the local changes and thereby synchronized. The network is divided into two-tier hierarchy (fig. 1). In the top-level overlay, CHORD concept is applied. And in the bottom-level hierarchy, the flow-sensors are clustered in single-connection manner. Flow-sensors communicate with the logical-sink. Sink that is part of a cluster virtually acts as a flow-sensor with very high-computational capabilities. This eliminates the burden of choosing or electing a cluster-head. This virtual flow-sensor can be thought as the cluster-head (one for each cluster). These virtual flow-sensors i.e. cluster-heads are organized in the top-tier overlay as CHORD. In fig. 1, there are three clusters that communicate with the logical-sink. And, for each cluster there is a virtual flow-sensor. A flow-sensor does not need to concern about the physical sink the communication takes place as all the physical sinks are synchronized and aware of any change inside the network.

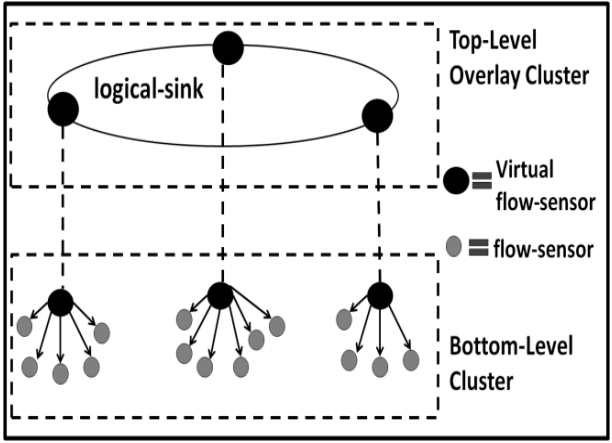

Fig. 1. The two-tier Network

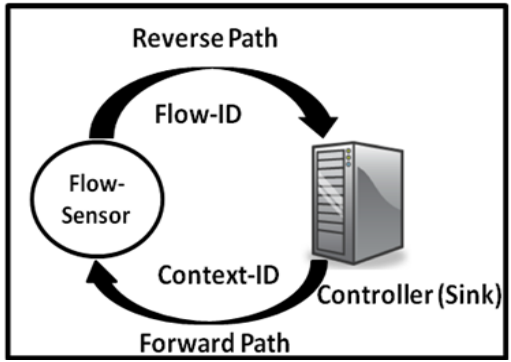

Fig. 2. Communication

## 4.2 Communication

The communication is shown in fig. 2 and is of three kinds: logicalsink-to-sensor, sensor-to-logicalsink and sink-to-sink. Logicalsink-to-sensor communication takes place in the forward path. This communication is straight forward in a sense that sink has better communication capabilities and can communicate with flow-sensors directly. For any exception, the communication can still take place through distributed collaboration. In the reverse path, sensors communicate with logical-sink via overlay hop [15]. Those sensors that are not part of a particular cluster collaborate with other sensors so that sensors can reach nearby logical-sink. Sink-to-sink (inside a HF network) communication can further be divided into two: physical and virtual sink-to-sink. The physical communication among sinks follows the same procedure as in HF. And, the virtual communication implies the communication between virtual flow-sensors and a CHORD top-level overlay is formed by the virtual flow-sensors. Hence, this communication follows the idea of CHORD.

## 4.3 Implementation

Both fixed and mobile flow-sensors have been assumed. Flow-sensors traffic are controlled and managed by logical-sink. The flow-sensor usually has flow-tables in the hardware layer [3]. Each flow-table contains flow-entries and an action for each flow-entry which decides flow routing. Each flow-entry has match-fields that define the flow, instructions correspond the way packets should be routed, and statistics takes care of packet updating. Packets from flow-sensors are matched in each flow-entry, instruction set defines the flow-ID if already not available, and statistics updates the packets. Statistics checks if the current packet matches the old packets, otherwise a new flow-ID is defined for any mismatch. Flow packets are then forwarded to the nearby physical sink in the reverse path. The flow packets include the flow-ID. The logical-sink maps flow-ID and returns the corresponding context-ID in the forward path, a sensor-ID is also returned to the flow-sensor if already not assigned. The sensor-ID is unique and unchanged for a flow-sensor. In case the context-ID is not available with the contacted physical sink, this sink contacts other physical sinks and the corresponding context-ID is returned. Search will follow the CHORD look-up mechanism. Viewed this way, the context-ID search will also follow the similar procedure. The logical-sink modifies the context flow-table with the context-ID along with sensor-ID and flow-ID. Logical-sink also updates the group table with the context-ID. By the mean time, other sinks get notified about all the changes in each sink and get updated thereby. In case the received flow-ID does not match any existing context-ID, then logical-sink defines a new context-ID. This context-ID is then published to other HF networks. When any sensor is interested in the context-ID in other network, then sensors subscribe to the context-ID. The algorithm for above is as follows:

- Flow-sensor match-fields define the flow and the action defines the flow-ID
- Flow-ID is sent to the nearby physical sink S1
- S1 resolves flow-ID and returns corresponding context-ID
- S1 returns the sensor-ID if already not assigned
- S1 forwards the request from flow-sensor to other physical sinks (S2, S3… Sn) if no match found for the request in S1
- If no context-ID found in the logical-sinks then a new context-ID is defined and published to other networks

- Logical-sink returns the context-ID to the requested flow-sensor
- Regular and context flow-tables are updated by the logical-sink
- Statistics check for new and old packet mismatch, new flow-ID is defined in case of any mismatch

### 4.4 Example Scenario

Fig. 3 shows an example of MATLAB implementation. There are 4 H-DHT HF net-works with 50 sensors. Some are fixed (16) and some are mobile (34). The sensors have been clustered based on context-similarity. Different cluster is represented by different color. As seen that sensors might be resided in different networks but they are logically clustered and belong to same context-ID. Each HF network has been facilitated by four sinks ('+' signs). Their positions are fixed and act as single logical-sink for single HF network. It is assumed that sinks are placed carefully so that all the flow-sensors are covered. This explains how logical clustering of sensors can be achieved.

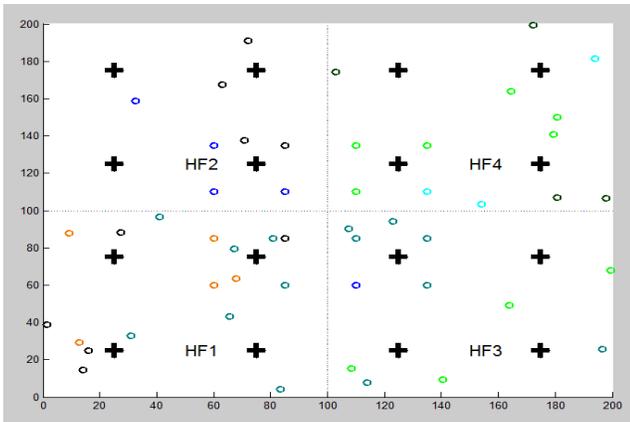

Fig. 3. An example scenario

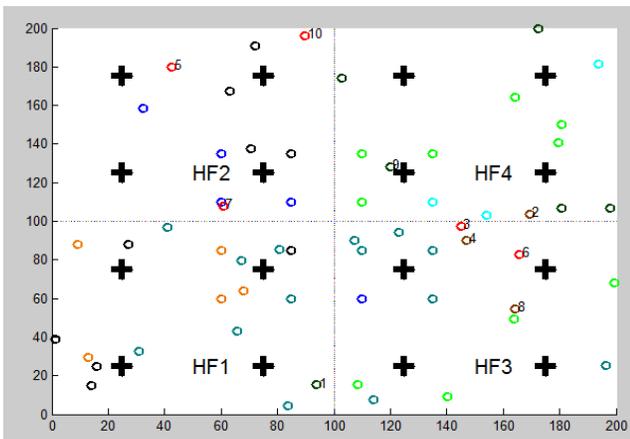

Fig. 4. Sensors joining

### 4.5 Sensors Joining

Fig. 4 shows 10 sensors (depicted by 1 to 10) joining. It can be seen that sensors (1, 9) have joined an existing cluster; while sensors (2, 4 and 8) and rest of the sensors have formed two new clusters respectively. These can be distinguished by different colors. When new sensors join the network, they send their flow-IDs to the nearby sinks. Context-IDs are shared by all the logical-sinks, and all the logical-sinks have the knowledge of existing context-IDs. Therefore, when sensors send their flow-IDs, then logical-sink checks the existing context-IDs. If match found, then new sensors are said to have subscribed to the existing context-ID. Otherwise, logical-sink defines new context-ID based on the received flow-IDs and context similarity. And, the sensors form new clusters.

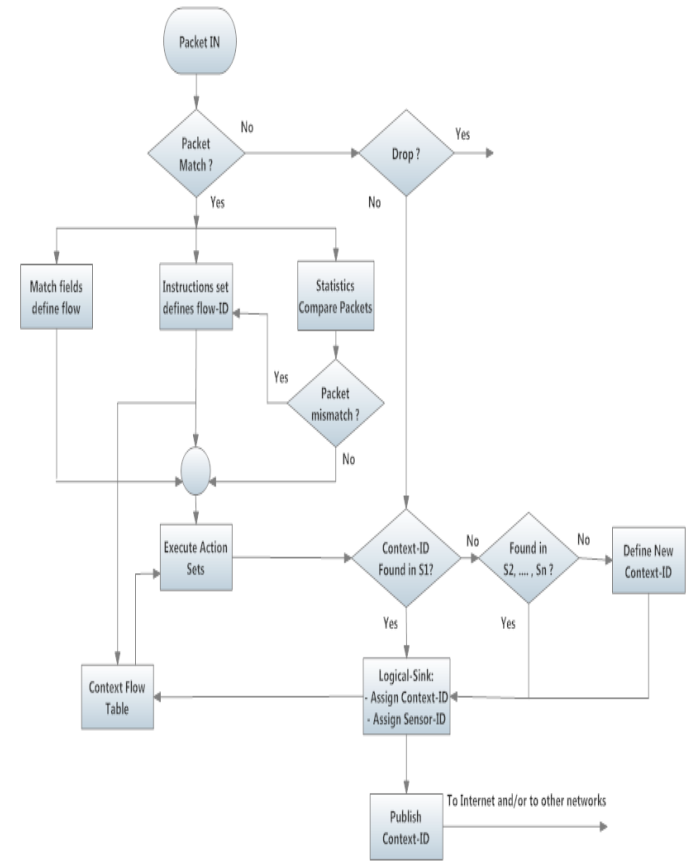

**Fig. 5.** Flow chart

### 5. Model Checking of the Concept

The combination of PROMELA and SPIN has been used for simulation and verification of system model in [17-19]. It provides versatility and is very useful for model checking. The combination has been used extensively for modeling and verifying communication protocols [3]. The proposed model of this paper has been examined using the PROMELA and SPIN combination. First, fig. 5 shows the flow chart of the proposed model. The explanation of the flow chart has been described already (see 4.3).

## 5.1 Context-ID Match Algorithm

The following algorithm defines the mechanism for communication between sensor nodes and sinks. The first process (proctype node) defines sensor nodes flow send and receive method, and the second process (proctype sink) defines the mechanism for logical sink.

```
/*Algorithm for context-ID definition
or matching*/

bool flow_id, sensor_id, context_id;
proctype node(chan in, out) {
#define node_add  /*define address of
the sensor node*/
int pkt; /*packet*/
bool chk;
xs src_node; /*send channel of source
node*/
xr sink_add; /*receive channel of
sink*/
in?input_port,dst_add; /*Channel sends
input port number and destination
address*/
if
:: (src_node == node_add && pkt! =Null)
-> out!pkt; goto pkt_match; /*if
address is authenticated and packet is
not empty, send packet to check for
packet matching*/
fi;
pkt_match: in?pkt
if
:: (chk = true) -> goto
pkt_send2flowtable; /*if packet is for
matching, send to flow table*/
:: (chk = false) -> goto pkt_drop;
/*check if packet is to be dropped*/
fi;
pkt_drop: in?pkt
if
:: (chk = true) -> skip; /*Packet is
dropped*/
:: (chk = false) -> pkt_send2sink;
/*Packet is forwarded to the nearby
physical sink*/
fi;
pkt_send2flowtable: in?pkt
if
:: (input_port == 1) ->
write(match_fields); /*If Packet is not
empty, update the match fields*/
:: (input_port == 2)
write(instructions_set); /*Update
instructions set and define the flow-
ID*/
:: (input_port == 3) write(stat);
/*Update Statistics- store packet
information*/
:: goto pkt_send2sink; /*Packet is
ready to be sent to the nearby sink*/
else -> skip; /*Drop Packet, if empty*/
fi;
pkt_send2sink: in?pkt
read(sink_add); /*Get the address of
the nearby sink*/
read(match_fields); /*Check the match
fields for flow*/
read(instructions_set); /*Check for
flow-ID*/
read(stat); /*Check for any packet
mismatch*/
if
:: (flow_id = true) -> goto
context_flowtable; /*If flow_id is
found then insert to the context flow
table*/
fi;
context_flowtable:
read(instructions_set);/*Update the
context flow table's flow-ID field*/
end; /*End the process*/
}
proctype sink(chan in,out) {
#define dst_add /*Define the current
sink address*/
if(dst_add == sink_add && pkt! =Null) -
> goto flow_match; /*If the sink
address is authenticated and packet is
not empty, then check for flow
matching*/
fi;
flow_match: in?pkt
read(match_fields); /*Check the match
fields for flow*/
read(instructions_set); /*Check for
flow-ID*/
read(stat); /*Check for any packet
mismatch*/
if
```

```
:: (flow_id = true) -> out!context_id
/*If flow-ID matches any existing
context, send the context-ID*/
:: (sensor_id = false) -
>write(sensor_id) /*If no sensor-ID is
assigned, assign the sensor-ID*/
:: out!sensor_id; /*Send the sensor-
ID*/
:: goto context_flowtable; /*Go to the
context flow table to update the table
fields*/
:: goto publish; /*Go to publish if
context-ID is ready to be published*/
::else goto sink_n; /*If flow does not
match any context-ID in the current
sink, go to other sinks*/
fi;
sink_n: in?flow_id
if
:: (flow_id = true) -> out!context_id
/*If flow-ID matches any existing
context, send the context-ID*/
:: goto context_flowtable; /*Go to the
context flow table to update the table
fields*/
:: goto publish; /*Go to publish if
context-ID ready to be published*/
::else write(context_id); /*If no
context-ID found for the flow, define a
new context-ID*/
fi;
context_flowtable:
in?flow_id,sensor_id,context_id
write(stat); /*Update the statistics
with IDs*/
publish: in?context_id
if
:: (context_id = false) ->
write(context_id); /*If context-ID is
not yet published, publish the ID*/
fi;
end; /*End the process*/
}
init { /*Initialize the processes*/
  chan send = [2] of {int, bool};
/*Send channel would carry two
different type of messages*/
  chan rcv = [2] of {int, bool};
/*Receive channel would carry two
different type of messages*/

  run node(send,rcv); /*run the node
process*/
  run sink(send,rcv); /*run the sink
process*/
}
```

## 6. Simulation Results

Table 1: Simulation parameters

| Parameter | Value |
|---|---|
| Number of Networks | 3 |
| Number of Nodes | 60 |
| Number of Groups | 3[*] |
| Nodes per Group | 9[*] |
| Packet Flow Rate (per second) | 8[*] |
| Packet Size | 512[*] bytes |
| Routing | Static |
| Propagation Path Loss Model | Fixed RSS Loss Model |
| Delay Model | Constant Speed Propagation Delay Model |
| Error Model | ns-3 YANS Error Model |
| Sensors Mobility Model | Random Walk 2d Mobility Model |
| Receiver Noise Factor | 10.25 dB |
| Received Signal Strength (RSS) | -95 dBm |
| Total Number of Transmitted Packets | 2000 |
| Physical Model | IEEE 802.11b |
| Data Rate | 1 Mbps |

[*] = varies in different simulations

A network has been designed and simulated in the ns-3 simulator. Simulation parameters are tabulated in table 1. The focus of this paper was not to verify the physical layer behaviors, hence the sensor node reachability, interferences, received signal strength (RSS), energy consumption, and signal-to-noise ratio impacts have not been explored. These are beyond the scope of this work. The focus largely lies on the behavior of the system with regard to real-time context sharing. Therefore, the simulation has been carried on constant values of RSS, receiver noise factor, etc.

6.1 Simulated Network

Fig. 6 shows the network that was designed and simulated in the ns-3. Although our proposal makes use of multiple distributed and synchronized OpenFlow controllers (logical-sink), but ns-3 as of now does not allow external controller for OpenFlow [20]. Hence, we stick to the current ns-3 implementation. As for H-DHT for sensor nodes and context-IDs management, this is also left for the upcoming paper as no working model of H-DHT is available right now in ns-3 [21]. In the designed network, there are three wireless sensor networks as seen in fig. 6. Sensor nodes in network 1 are fixed while sensor nodes in both network 2 and 3 are mobile (randomly moving). Each network has one gateway and gateways are connected by the OpenFlow controller. Each network has 20 sensor nodes. Other parameter values can be found in table no. 1.

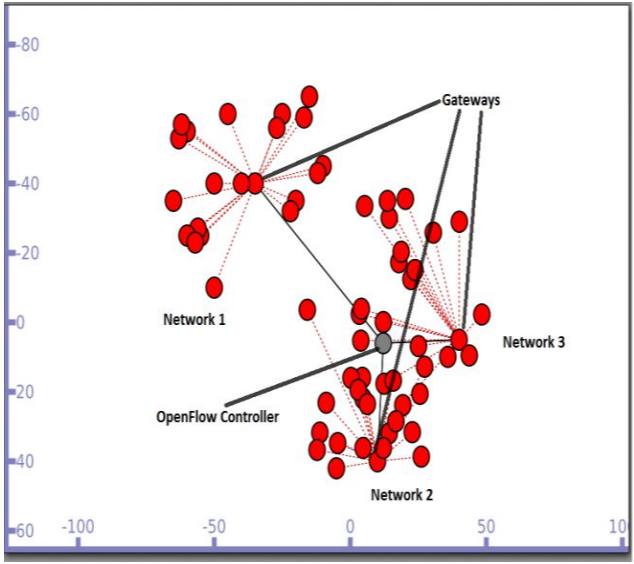

**Fig. 6.** Simulated Network

## 6.2 Performance Measurement

In this section, performance measurement for various scenarios has been presented. For the evaluation, reliability, scalability and reachability metrics have been chosen and the proposed approach has been highlighted with respect to these metrics. The importance of reliability and scalability has been suggested by the earlier researches. As for reachability, we believe that packet reachability would be an important performance metrics in the real-time context sharing e.g. in urban event detections.

### 6.2.1 Effect of Varying Flow-Rate

Firstly, the performance has been measured for different flow rate i.e. number of packet per second (p/s). The number of node per group and the group size has been kept unchanged for this particular evaluation. There are total 3 groups for this scenario and each group has 9 nodes. The packet size for this scenario has been kept to be 512 bytes.

*Mean Delay Performance*

Fig. 7 shows mean delay performance for the simulated network of each group for different flow of packet. Packet flow varies between 6 and 11 p/s. X-axis shows the packet flow rate and y-axis shows the mean delay of each group. It can be seen from the figure that at the start, i.e. for packet flow of 6 p/s, each group more or less demonstrates similar results with respect to delay. All groups qualitatively demonstrate similar performance for packet flow rate up to 10 p/s. While the packet flow is increased to 11 p/s, all groups show increase in the delay for 11 p/s. It can be seen that delay is increased with the increase in the packet flow rate, however, it does not incur high increase up to 10 p/s. Group 1, 2 and 3 mean delay increase by 0.3172s, 0.2629s and 0.2166s respectively for 11 p/s i.e. 83 % increase in the packet flow rate.

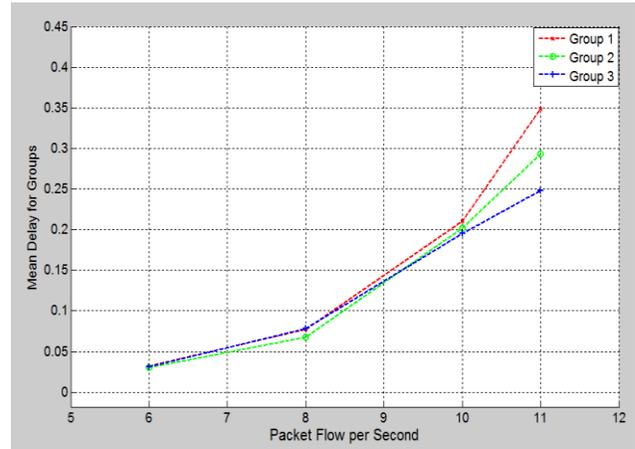

Fig. 7. Mean Delay Performance

*Mean Jitter Performance*

Mean jitter performance for each group for different flow of packet can be seen in fig. 8. Mean jitter performance demonstrates similar pattern like mean delay. At the beginning, all groups show similar jitter performance. Like mean delay of the packet, jitter does not encounter a performance degradation when packet flow is increased. Jitter for 11 p/s increases by only 0.0347s, 0.0327s and 0.0313s for group 1, 2 and 3 respectively compared to 6 p/s. The increase is very minimal. Therefore, from fig. 7 & 8 it can be concluded that packet delivery is reliable with minimum delay and jitter.

Packet delivery with minimum delay and jitter is very significant issue in crowdsourcing paradigm and for any urban event detection. When crowd generate data, it should be delivered swiftly. Service requesters would want to access data in the shortest possible time. Moreover, reliability of packet delivery in real-time context sharing largely depends on how quickly service is delivered. Another important characteristic scalability can be seen from fig. 10 &11. As for the packet flow rate, packet size plays an important role defining flow rate. Fig. 16 shows the impact of changing packet size.

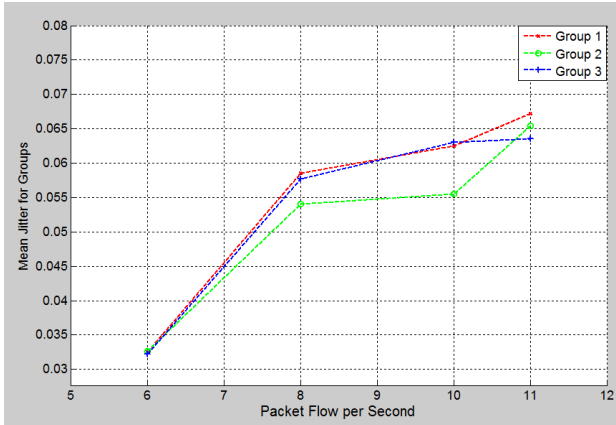

Fig. 8. Mean jitter performance

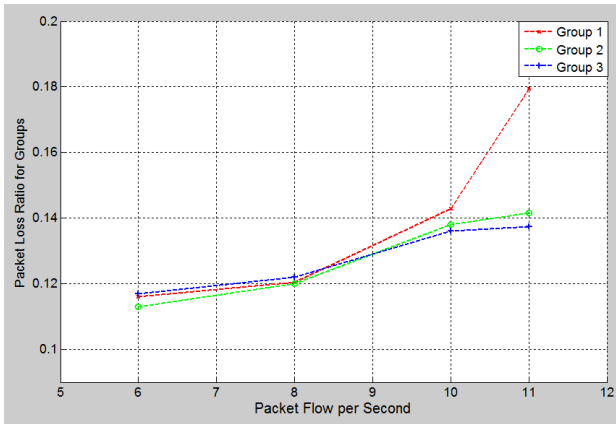

Fig. 9. Packet Loss Ratio

*Packet Loss Ratio*

Reachability of packet principally depends on number of packet loss. Fig. 9 shows packet loss ratio of each group. Although each group demonstrates similar pattern in packet loss for packet flow rate up to 10 p/s. Group 1 shows a rise in the packet loss for flow rate 11 p/s. Packet loss ratio increases with the increase in the flow rate. For the packet size of 512 bytes, packet loss ratio does not incur a high fluctuation for flow rate up to 10 p/s. This assures high reachability of packet. Group 1, 2 and 3 incurs packet loss ratio increase by 0.0635, 0.0285 and 0.0205 respectively compared to flow rate of 6 p/s. In terms of percentage, the increase is 55 %, 25% and 18% compared to flow rate of 6 p/s. This increase of packet loss is for 83 % increase in the flow rate. From this, it is clear that our proposal assures rich packet reachability. In the real-time urban event detection, this high reachability of packet would be very beneficial. This will ensure sharing rich amount of urban event detections.

### 6.2.2 Effect of Increasing Nodes per Group and Group Size

In the previous section, reliability and reachability have been discussed for variant flow of the packets. Although in the urban event detection packet flow rate would always be fluctuating, however, at the same time the participants in data acquisition i.e. sensor nodes in this case would also vary. This implies that different context would be generated which leads to different clustering of contexts i.e. group of data. Scalability becomes a significant consideration with respect to increasing number of nodes and groups for real-time context management. Here, effect of increasing nodes per group and group size is discussed.

*Mean Delay Performance*

Fig. 10 shows the mean delay performance for variant number of nodes per group. As seen earlier that performance degrades from 11 p/s, and packet flow rate has been kept constant at 10 p/s for this evaluation. As expected, delay increases with the increase in the node per group. If the node per group is doubled then mean delay increases by 17 %, 18 % and 15 % respectively for group 1, 2 and 3. And, if the node per group is tripled i.e. increase by 200% then group 1, 2 and 3 incurs mean delay increase by 25 %, 20 % and 22 % respectively. This clearly shows that the proposed concept scales well for increased number of node per group. Figure 11 shows impact of increasing the number of groups. The figure illustrates only results for group 1 and node per group has been kept steady (9 node per group). It has been evaluated for different packet flow rate. As seen from figure 11, mean delay increases nominally with increase in size of the groups. For instance, for the packet flow rate of 5 p/s, group size of 6 incurs 6 % delay increase compared to group size of 3. For the packet flow rate of 8 p/s, the delay increased to 19 %. It is seen that only 13% delay fluctuated when flow rate is increased by 60 % and group size is doubled. However, it is observed that for group size of 6 with flow rate 9 p/s, mean delay decreased. This is due to the fact that packet loss for this scenario is higher due to probable wireless interferences and random nodes' movement. It can be concluded that the proposed concept provides scalability for delay in terms of increased node per group and increased group size.

*Mean Jitter Performance*

Mean jitter performance for scalability is shown in next two figures. Fig. 12 shows the effect of changing node per group on jitter. As was the case with the mean delay, jitter also understandably increases with raise in the node per group. If node per group is doubled (100 % increase) then jitter increases by 33 %, 35 % and 33% for group 1, 2 and 3 respectively compared to 6 nodes per group. And if the node per group is increased by 200% then group 1, 2 and 3 encounter jitter increase by 41 %, 40 % and 35 % respectively compared to 6 nodes per group. This clearly shows the proposed concept scales well in terms of jitter too. Figure 13 further shows mean jitter performance while keeping node per group constant (9 node per group), and varying the size of the group and flow rate. The figure evidently demonstrates that jitter fluctuates nominally for the aforementioned scenario. For the flow rate of 8 p/s, jitter demonstrates only 44 % fluctuation for 100 % increase in the group number. Due to packet loss it is observed that jitter decreases for flow rate of 9 p/s. This packet loss depends on the flow rate and packet size. Figure 16 clarifies effect of packet size variation. This low jitter fluctuation will particularly provide advantage in crowdsourcing paradigm, when there would be different number of clusters of context generated by crowd. Different clusters of context imply different types of urban events detection. Therefore, our proposal can cope in terms of reliability with reliable real-time context sharing for changeable number of nodes and clusters.

*Packet Loss Ratio*

Along with the reliable context delivery it is also imperative that context delivery ratio is high and scalable at the same time. Figure 14 shows the packet loss ratio performance for variant number of node per group. As seen from the figure 14, when node per group is increased by 100% then group 1, 2 and 3 respectively have packet loss increased by 22 %, 22 % and 41 %. And, if the node per group is increased by 200% i.e. to the full capacity of the designed network, packet loss ratio increases by 71 %, 48 % and 102 % for group 1, 2 and 3 respectively. Group 1 and 3 exhibited higher packet loss compared to group 2. Also seen from the figure, for packet flow rate of 9 p/s, group 1 has a leap in the packet loss for group size of 5 and 6. This higher ratio is due to random movement of the nodes and wireless interferences. Compensating these effects is beyond the scope of this work. It can further be seen from figure 15 that group size of 6 exhibits packet loss ratio degradation only by 26 % and 33 % compared to group size of 3 (100 % increase in the group size) for flow rate of 5 p/s and 8 p/s respectively. This confirms that the proposed idea is scalable for packet reachability as well.

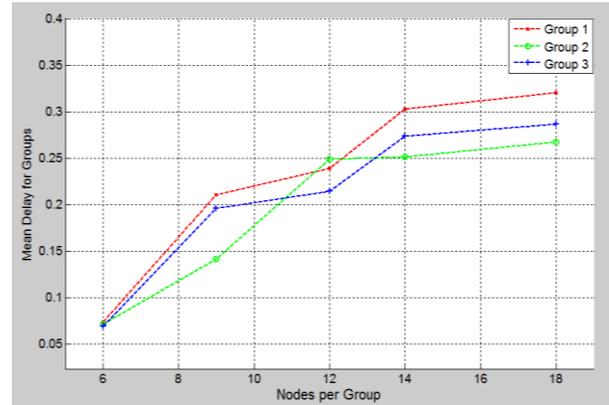

Fig. 10. Mean delay performance for different number of nodes per group

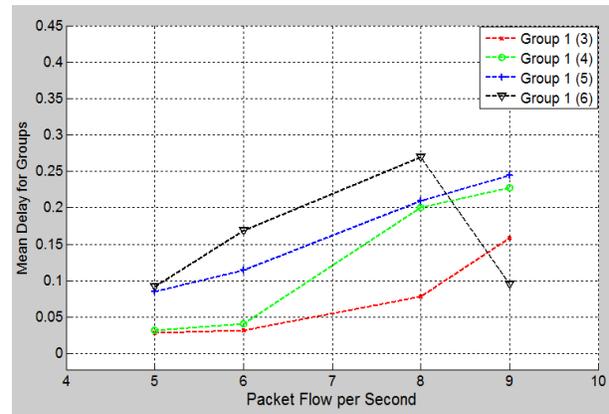

Fig. 11. Mean delay for variant size of groups

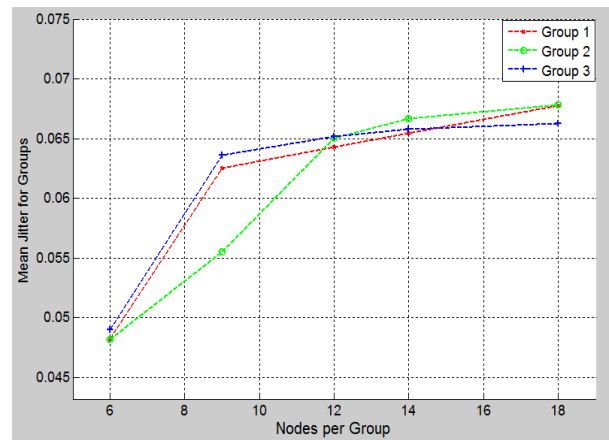

Fig. 12. Mean jitter for alternate number of node per group

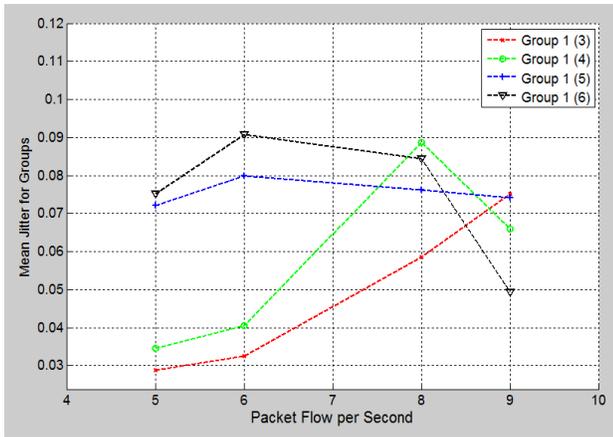

Fig. 13. Mean jitter for variant size of groups

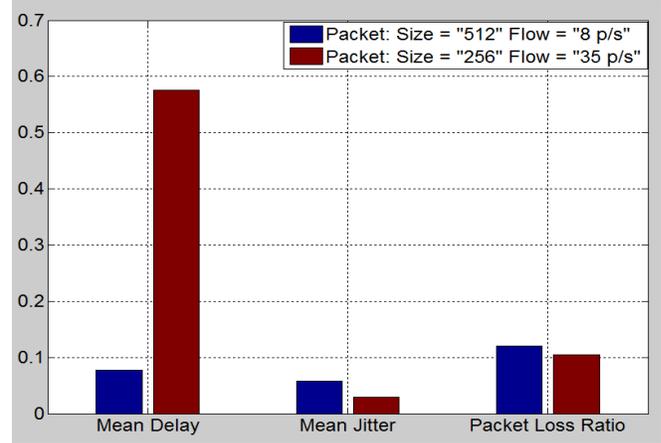

Fig. 16. Packet size impact on packet flow rate

### 6.2.3 Different Packet Size

The above results have been evaluated for a particular packet size of 512 bytes. It is clear that for this packet size, performance metrics shows better results if packet flow rate is below or equal to 10 p/s. Now fig. 16 shows how packet size affects the packet flow rate. It can be seen that if the packet size is halved then packet flow rate increases by 338 %. However, mean delay also increases by 642 %. As for mean jitter and packet loss ratio, these metrics decreased by 98 % and 15 % respectively. From this it can be concluded that in the crowdsourcing paradigm or in the urban event detection, if the generated data i.e. context is small then packet flow rate will be high. However, this might result in high delay but jitter and packet loss ratio would be lower. Therefore, with high packet flow rate, the idea can scale well for jitter and packet loss ratio but delay performance might degrade.

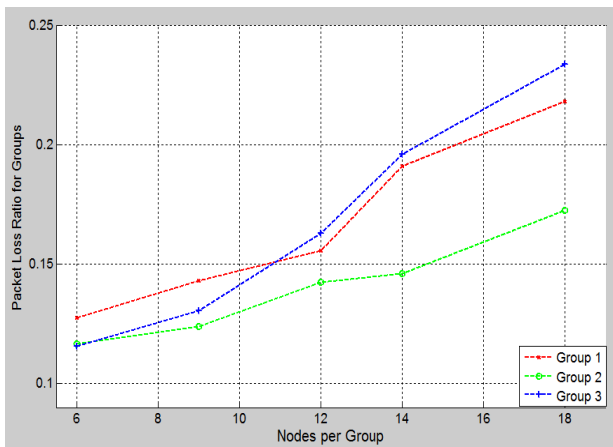

Fig. 14. Packet loss ratio for different number of node per group

## 7. Use Cases of the Concept

The proposed approach would be useful for heterogeneous interoperability of physical objects, thereby heterogeneous contexts. In our opinion, this logical-clustering will be advantageous to many sensor network applications; for example, medical science, agriculture system, security surveillance, disaster management etc. Two probable scenarios are portrayed below.

### 7.1 Animal Tracking

The use of WSN for animal tracking is gaining tremendous attention recently [22]. The animal tracking can be further divided into two: wildlife and farming monitoring. Our proposed approach can be applied in both of them. One probable application, for example, farmers can optimize their business by means of WSN for feeding and growing conditions of the animals [22]. This will provide benefit for monitoring meat, milk production and to observe how good animal racing results. Hence, animal tracking would be easier by applying our proposed logical-clustering approach. This implies that animals that produce similar desired context would be clustered together. In a large farm, it is often difficult to manage the animals efficiently; it would require incredible manpower to monitor all the animals. Therefore, animals' location and conditions can be monitored by clustering. The farmers can find the groups of animal ready for meat and milk production through the context-ID. This will reduce human labor to find out the animals for the above mentioned purposes.

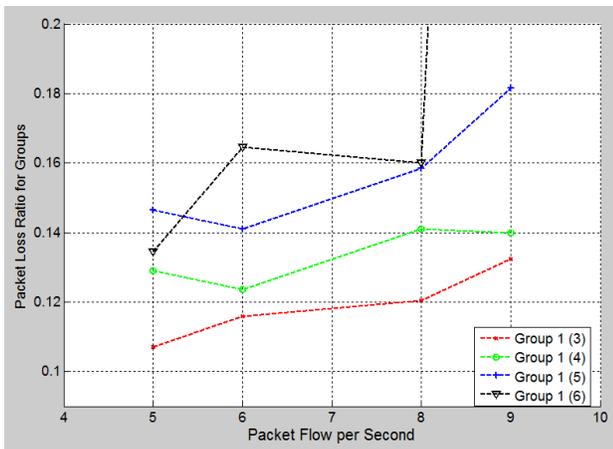

Fig. 15. Packet loss ratio for variant number of groups

7.2 Medical Healthcare

The approach can be applied in the medical healthcare too. One possible application scenario can be that medical researchers can conduct a research in real-time on a recently spread disease from remote places and provide prompt solutions simultaneously. Normally in medical healthcare, patients are outfitted with wireless wearable sensors [23]. If there is any outbreak of a disease, then people can be outfitted with wearable sensors. Medical team can observe the severity of the disease by means of clustering different symptoms and different level of patients. For example, patients might not have same level of severity and they would need different level of attentions. Hence, medical assistance can be provided faster and efficiently by clustering patients based on the different context of severity. Patients that show similar symptoms would be clustered together and would have same context-ID. This will eliminate burden of individual attention for a patient. Thus medical services can be maximized. Moreover, if the situation gets worse and out of control then medical personnel can seek help from other specialists in the respective field from remote places. In the traditional way, this can be done by gathering data from people and then forwarding to others. However with our proposed logical-clustering, medical researchers from distant places can instantly access the data by subscribing to the context-ID.

However, our proposal is not limited to these scenarios rather this shows two of the many possible solutions our proposal can offer.

## 8. Conclusions

Real-time context sharing would be an important challenge in state-of-the-art ubiquitous computing. The enormous data that are expected to be generated by the billions of sensors would require efficient management. These huge heterogeneous data would need to be processed reliably, and reachability should also be assured to take its full advantage. Location agnostic clustering of flow-sensors i.e. logical clustering is one of the possible solutions for efficient context management. In this paper, performance of logical-clustering in terms of delay, jitter and packet loss ratio has been examined and backed up by ns-3 simulations. These parameters have been evaluated for different scenarios such as: variant packet flow rate, different number of node per group and different group size.

The results suggest that the proposal is reliable and scalable. For a 200 % increase in the node per cluster, delay increases by around 20 %. For the same increased node per cluster, latency demonstrates around 40 % increase. Delay and latency exhibit 13 % and 44 % fluctuation respectively when the cluster size is doubled. This clearly illustrates efficiency of reliability and scalability of the proposed concept. Furthermore, the proposed approach shows rich packet reachability. Only 18 % increase in the packet loss for a flow rate increase of 83 %, packet loss increases by merely 44 % for 200 % increase in the node per cluster. Moreover, packet loss demonstrates no more than 33 % increase for 100 % increase in the cluster size for high flow rate.

The proposed approach in particular can perform more efficiently for smaller packet size as suggested by the result. Flow rate increased by 338 % when packet size is halved. Latency and packet loss ratio further decreased by 98 % and 15 % respectively. Two tangible use cases have also been portrayed. Therefore, our proposed idea will be of great interest for the future Networked Society where instantaneous and reliable accesses to context are two of the big challenges. Our approach can adopt quickly and share real-time data reliably to the service requestors. The vision of detecting any urban event via crowdsourcing paradigm will be made easier through the adoption of our proposal.

However, the approach can perform better than the results obtained in this paper through real implementation of logical-sink and H-DHTs in ns-3. The logical-sink would outperform the current packet reachability; reliability and response time would also be minimal. Future work includes designing the system with logical-sink and inclusion of H-DHTs. An investigation into routing protocol for the logical-clustering would also be explored.